\documentclass[a4paper,twocolumn]{article}
\usepackage{amsfonts,amsmath}
\newcommand{\mS}{{\mathcal S}}
\newcommand{\mO}{{\mathcal O}}

\begin{document}

%%%%%%%%%%%%%%%%%%%%%%%%%%%%%%%%%%%%%%%%%%%%%%%%%%%%%%%%%%%%%%%%%%%%%%%%%%%%%%

\title{On the instabilities of  the static, spherically symmetric SU(2) Einstein-Yang-Mills-Dilaton solitons and black holes}

\author{Walter H. Aschbacher\footnote{aschbacher@ma.tum.de}\\
\\
Technische Universit\"at M\"unchen \\
Zentrum Mathematik, M5\\
 85747 Garching, Germany}

\date{}
\maketitle

%%%%%%%%%%%%%%%%%%%%%%%%%%%%%%%%%%%%%%%%%%%%%%%%%%%%%%%%%%%%%%%%%%%%%%%%%%%%%%
\begin{abstract}

We prove that the number of sphaleron instabilities of the $n$-th SU(2) Einstein-Yang-Mills-Dilaton soliton and  black hole equals $n$.
 
\end{abstract}

%%%%%%%%%%%%%%%%%%%%%%%%%%%%%%%%%%%%%%%%%%%%%%%%%%%%%%%%%%%%%%%%%%%%%%
\section{Introduction}

The discovery of a countable family of solitons \cite{BK88} and black holes \cite{VG89} in static, spherically symmetric SU(2) Einstein-Yang-Mills (EYM) theory has triggered an ongoing interest in the study of matter models of this kind. The answer to the central question about the (linear) stability of these solutions turned out to be negative though \cite{SZ}: The spherically symmetric perturbations which decouple into an even and an odd parity sector both contain negative modes \cite{LM95,VBLS95,V96}. Whereas the former are called ``gravitational'' (not having  flat space-time analogues), the latter allow for a ``sphaleron-like'' interpretation, cf. \cite{LM95,VBLS95,V96} and references therein. Numerical studies in \cite{SZ} showed that the $n$-th soliton or black hole, characterized by the number $n$ of nodes of the gauge potential, has exactly $n$ negative gravitational modes. In the sphaleron sector,  this has been shown numerically in \cite{LM95}  whereas in \cite{VBLS95,V96} the following strategy has been used for an analytical proof: The arising system of coupled differential equations for the perturbation amplitudes is cast into the form of a Schr\"odinger equation. Its formal supersymmetric partner Hamiltonian has a zero energy state which depends on an auxiliary function whose properties, due to the residual gauge freedom of the perturbation equations, can be chosen such that the roots of the zero mode are the roots of the background gauge potential. Since the supersymmetry transformation is reversible for strictly negative energies, the number of negative modes thus equals $n$. 

It is important to ask what role these EYM solitons and black holes play in string theory. One of the most widely studied stringy corrections stems from the superstring effective action at low energies which supplements the EYM theory, among other scalar fields, by the massless neutral dilaton with exponential coupling to matter (Kaluza-Klein theories and cosmological models also suggest such a supplementation). This stringy generalization, the so-called Einstein-Yang-Mills-Dilaton (EYMD) theory, is specified by the action
\begin{eqnarray}
\label{action}
\mS=\frac{1}{4\pi}\int\!\!\sqrt{-g}\,d^4x\,\left(-\frac{1}{4G}R+\frac{1}{2}(\partial\varphi)^2-\frac{e^{2\kappa\varphi}}{4\lambda^2}F^2\right),
\end{eqnarray}
where $R$ is the scalar curvature, $G$ Newton's constant, $\phi$ the dilaton field with coupling constant $\kappa$, and $F$ the Yang-Mills field with coupling constant $\lambda$. After a suitable rescaling (cf. after \eqref{S3}), the theory is parametrized by 
\begin{eqnarray}
\label{gamma}
\gamma=\frac{\kappa}{\sqrt{G}}.
\end{eqnarray}
The EYM and the Yang-Mills-Dilaton theory, both embedded in the EYMD theory, are recovered from the latter in the limits $\gamma\to 0$ and $\gamma\to \infty$, respectively. As for these extreme cases \cite{BK88,LM92},  due to the breaking of scale invariance of the Yang-Mills  action, a discrete familiy of static, spherically symmetric soliton and black hole solutions has been found in \cite{LM93}  for the intermediate values $0<\gamma<\infty$ (for the extreme cases, rigorous existence proofs and classifications of the solutions are given in \cite{SW93}).  In this paper, we study the instabilities of these generalized solutions in the sphaleron sector. Using the strategy described above, we prove that the number of instabilities of the $n$-the soliton or black hole equals the number $n$ of nodes of the gauge potential.

%%%%%%%%%%%%%%%%%%%%%%%%%%%%%%%%%%%%%%%%%%%%%%%%%%%%%%%%%%%%%%%%%%%%%%
\section{The number of instabilities}
\label{section:instabilities}

In order to compute the action $\mS$ from \eqref{action} for the spherically symmetric EYMD theory and, eventually, to derive the corresponding background and pulsation equations, we parametrize the spherically symmetric gravitational field by Schwarzschild coordinates,
\begin{eqnarray}
\label{ds2}
ds^2=NS^2\,dt^2-\frac{dr^2}{N}-r^2(d\theta^2+\sin^2\theta\,d\phi^2),
\end{eqnarray}
and the su(2)-valued gauge potential by
\begin{eqnarray}
\label{A}
A&=&a_0\tau_r\,dt+a_1\tau_r\,dr+(w-1)(\tau_\phi\,d\theta-\tau_\theta\sin\theta\,d\phi)\nonumber\\
&&+{\tilde w}(\tau_\theta\,d\theta+\tau_\phi\sin\theta\,d\phi),
\end{eqnarray}
cf. \cite{BBMSV94}. The metric coefficients $N$, $S$ and the gauge amplitudes $a_0$, $a_1$, $w$, ${\tilde w}$ depend on $r$ and $t$ only. The spherical generators $\tau_j$ of su(2), $j=r,\theta,\phi$, are  defined by $\tau_j=\tau\cdot e_j$, where $\tau=\sigma/(2i)$ with the usual Pauli matrices $\sigma=(\sigma_1,\sigma_2,\sigma_3)$, and $e_r$, $e_\theta=\partial_\theta e_r$, and $e_\phi=\partial_\phi e_r/\sin\theta$ denote the unit vectors in the $r$, $\theta$, and $\phi$ directions, respectively.\\
Plugging the parametrizations \eqref{ds2},\eqref{A} into \eqref{action}, we compute the scalar curvature $R$ using the Cartan structure equations and the gauge field strength by $F=dA+[A,A]/2$. Then, the  action becomes
\begin{eqnarray}
\label{action-densities}
\mS=\frac{1}{2}\int\!dr dt \,\left(\frac{1}{G}\,\mS_1+\mS_2+\frac{e^{2\kappa\varphi}}{\lambda^2}\,\mS_{3}\right),
\end{eqnarray}
where the densities are given by the expressions
\begin{eqnarray}
\label{S1}
\mS_1&=&-S\,(N+rN'-1),\\
\label{S2}
\mS_2&=&r^2\,\left(\frac{{\dot \varphi}^2}{NS}-NS\varphi'^2\right),\\
\label{S3}
\mS_{3}&=&\frac{r^2}{2S}\,({\dot a}_1-a'_0)^2\nonumber\\
&&+\frac{1}{NS}\,(({\dot w}+a_0{\tilde w})^2+({\dot {\tilde w}}-a_0w)^2)\nonumber\\
&&-NS\,((w'+a_1{\tilde w})^2+({\tilde w}'-a_1w)^2)\nonumber\\
&&-\frac{S}{2r^2}\,(w^2+{\tilde w}^2-1)^2.
\end{eqnarray}
For the length $\delta=\sqrt{G}/\lambda$, the rescaling of the form $r\mapsto\delta r$, $t\mapsto t/\delta$, $S\mapsto\delta^2 S$, $\varphi\mapsto \varphi/(\lambda\delta)$, $a_0\mapsto\delta a_0$, $a_1\mapsto a_1/\delta$\, ($N$, $w$, ${\tilde w}$ unchanged) allows to set $\kappa=\gamma$ from \eqref{gamma} and $G=\lambda=1$ to which choice we will stick in the following.

In the static case, and with the magnetic ansatz $a_0=a_1={\tilde w}=0$, the stationarity of the action \eqref{action-densities} with \eqref{S1},\eqref{S2}, \eqref{S3}, leads to the background Euler-Lagrange equations for the fields $N,S,w$, and $\varphi$, respectively,
\begin{eqnarray}
\label{bg_1}
S'&=&S\left(r\varphi'^2+\frac{2}{r}e^{2\gamma \varphi}w'^2\right),\\
\label{bg_2}
(rN)'&=&1-r^2N\varphi'^2-U\\
\label{bg_3}
(e^{2\gamma \varphi}NSw')'&=&e^{2\gamma \varphi}\,S\,\frac{w(w^2-1)}{r^2},\\
\label{bg_4}
(r^2 NS\varphi')'&=&\gamma SU,
\end{eqnarray}
where we define
\begin{eqnarray}
\label{U}
U=2e^{2\gamma \varphi}\left(\!Nw'^2\!+\!\frac{(w^2-1)^2}{2r^2}\right).
\end{eqnarray}

Since we are interested in a linear stability analysis for small spherically symmetric perturbations of the background soliton and black hole solutions of \eqref{bg_1}--\eqref{U}, we determine the pulsation equations for the perturbing fields $\delta N$, $\delta S$, $\delta a_0$, $\delta a_1$, $\delta w$, $\delta {\tilde w}$, and $\delta \varphi$ by linearizing the general Euler-Lagrange equations around these background solutions, cf. \cite{BHS96} for a general procedure. As for the EYM case mentioned in the introduction, the gravitational  modes $\delta N$, $\delta S$, $\delta w$, $\delta\varphi$ and the sphaleron modes $\delta a_0$, $\delta a_1$, $\delta {\tilde w}$ decouple in the EYMD system in the magnetic ansatz $a_0=a_1=\tilde{w}=0$. With the temporal gauge $\delta a_0=0$, the pulsation equations for the sphaleron modes which we are interested in become
\begin{eqnarray}
\label{a1}
\ddot{\delta a_1}&=&\frac{2NS^2}{r^2}\,\left(w\,\delta\tilde{w}'-w'\delta\tilde{w}-w^2 \delta a_1\right),\\
\label{ww}
\ddot{\delta {\tilde w}}&=&e^{-2\gamma \varphi} NS\,\left(e^{2\gamma \varphi} NS\left(\delta{\tilde w}'-w\delta a_1\right)\right)'\nonumber\\
&&-N^2S^2w'\delta a_1-\frac{NS^2}{r^2}\,\left(w^2-1\right)\,\delta {\tilde w},\\
\label{a0}
w \,\dot{\delta\tilde{w}}&=&e^{-2\gamma \varphi}NS\,\left(\frac{r^2}{2S}\,e^{2\gamma \varphi}\,\dot{\delta a_1}\right)'.
\end{eqnarray}
To simplify the form of these equations, we introduce the tortoise coordinate $\rho$ defined by 
\begin{eqnarray}
\label{def:tortoise}
\frac{d\rho}{dr}=\frac{1}{NS},
\end{eqnarray}
with $\rho(0)=0$ for the solitons, and $\rho=-\infty$ at the horizon of the black holes (from here on to the end of this section, the prime will denote the derivative with respect to $\rho$). Plugging the perturbations 
\begin{eqnarray}
\label{perturb}
\delta a_1=\alpha(\rho)\,e^{i\omega t},\quad \delta\tilde{w}=\beta(\rho)\,e^{i\omega t}
\end{eqnarray}
into the linearized dynamical equations \eqref{a1}, \eqref{ww}, \eqref{a0},  we are left with a coupled system of ordinary linear differential equations for the gauge amplitudes $\alpha$ and $\beta$ consisting of 
\begin{eqnarray}
\label{1st_ord}
\frac{1}{2}\omega^2\alpha&=&\frac{S}{r^2}(w'\beta-w\beta')+\frac{NS^2}{r^2}w^2\alpha,\\
\label{2nd_ord}
\omega^2e^{2\gamma \varphi} \beta&=&-(e^{2\gamma \varphi} \beta')'+e^{2\gamma \varphi}\frac{NS^2}{r^2}(w^2-1) \beta\nonumber\\
&&+e^{2\gamma \varphi} NS w' \alpha+(e^{2\gamma \varphi} NSw \alpha)',
\end{eqnarray}
and the Gauss constraint
\begin{eqnarray}
\label{gauss1}
\omega \left(\left(e^{2\gamma \varphi} \frac{r^2}{2S}\alpha\right)'-e^{2\gamma \varphi} w\,\beta\right)=0.
\end{eqnarray}
To rewrite the equations \eqref{1st_ord}, \eqref{2nd_ord}, \eqref{gauss1} in a more compact form, we define  modified gauge amplitudes $x$, $y$,
\begin{eqnarray}
\label{def:x,y}
x=e^{2\gamma \varphi} \frac{r^2}{2S}\,\alpha,\quad y=e^{\gamma \varphi} \beta,
\end{eqnarray}
and modified background fields $f$, $g$,
\begin{eqnarray}
\label{def:f,g}
f=e^{\gamma \varphi} w,\quad g=e^{-2\gamma \varphi} \frac{2NS^2}{r^2}.
\end{eqnarray}
Now, note that, if we multiply the right hand side of equation \eqref{1st_ord} by $e^{2\gamma \varphi} r^2/S$, take the derivative of the resulting product, and make use of the background equation \eqref{bg_3} in the  coordinate $\rho$, 
\begin{eqnarray}
\label{bg_3tortoise}
(e^{2\gamma \varphi}w')'&=&e^{2\gamma \varphi}\,\frac{NS^2}{r^2}\,w(w^2-1),
\end{eqnarray}
we get the right hand side of equation \eqref{2nd_ord} multiplied by $w$. Hence, using the definitions \eqref{def:x,y}, \eqref{def:f,g}, we can rewrite equation \eqref{1st_ord} as $f^2gx+f'y-fy'=\omega^2 x$, equation \eqref{2nd_ord} as $(f^2gx+f'y-fy')'=\omega^2 fy$, and the Gauss contraint \eqref{gauss1} as $\omega(x'-fy)=0$. If $\omega\neq 0$, the second equation is a consequence of the first and the third one. In this case, we are left with the system
\begin{eqnarray}
\label{eq_1'}
fy'-f'y&=&\left(f^2g-\omega^2\right) x,\\
\label{gauss'}
x'&=&fy.
\end{eqnarray}
In the case  $\omega=0$, the Gauss constraint is void and we have to deal with the single pulsation equation \eqref{eq_1'} which is invariant under the gauge transformation $x\mapsto x+h'/g$ and $y\mapsto y+hf$, where $h$ is an arbitrary function. Hence, 
\begin{eqnarray}
\label{gauge-modes}
x=\frac{h'}{g},\quad y=hf
\end{eqnarray}
solve \eqref{eq_1'}. In the following, we keep the Gauss constraint \eqref{gauss'} also for the case $\omega=0$ where it plays the role of a gauge fixing condition (also called the strong Gauss constraint): The remaining gauge invariance in \eqref{gauge-modes} is broken in the sense that $h$ has to fulfill the linear second order differential equation 
\begin{eqnarray}
\label{gauge-fixing}
\left(\frac{h'}{g}\right)'=f^2h.
\end{eqnarray}
Now, as in \cite{VBLS95,V96}, to determine the number of instabilities, we cast the system \eqref{eq_1'},\eqref{gauss'} into the form of a formal radial Schr\"odinger equation for the function $\phi=x/f$ by eliminating the amplitude $y$ with the help of the strong Gauss constraint \eqref{gauss'}. We then get 
\begin{eqnarray}
\label{H0}
H^0\phi=\omega^2\phi,
\end{eqnarray}
where the potential $V^0$ in the Hamiltonian $H^0=-d^2/d\rho^2+V^0$ is given by
\begin{eqnarray}
\label{V0}
V^0&=&2\,\left(\frac{f'}{f}\right)^2-\frac{f''}{f}+f^2g.
\end{eqnarray}
Since $V^0$ has singularities at the roots of the gauge amplitude $w=e^{-\gamma\varphi}f$, we try to construct the supersymmetric partner Hamiltonian $H^1$ of $H^0$ whose potential $V^1$ is such that the number of bound states of $H^1$ can easily be determined. We proceed as it has been described in the introduction. Due to the residual gauge invariance \eqref{gauge-modes},\eqref{gauge-fixing},  the Hamiltonian $H^0$ has a zero mode $\phi_0=h'/(fg)$. This zero mode allows for the facorization $H^0=Q^+Q^-$ where the charges $Q^\pm=\mp d/d\rho+\sigma$ are expressed through the superpotential $\sigma=-\phi_0'/\phi_0$ which solves  the Riccati equation $V^0=\sigma^2-\sigma'$. Moreover, the potential of the supersymmetric partner $H^1=Q^-Q^+=-d^2/d\rho^2+V^1$ is given by $V^1=\sigma^2+\sigma'$. Using  the definition $z=-gh/h'$ and the strong Gauss constraint \eqref{gauge-fixing}, the superpotential takes the form $\sigma=f'/f+f^2z$. Therefore, since $z$ solves the Riccati equation $z'=f^2z^2-g$, the first term on the right hand side  of \eqref{V0} cancels. Hence, $V^1$ has the form 
\begin{eqnarray*}
V^1=\frac{f''}{f}+f^2g+2(f^2z)'.
\end{eqnarray*}
Moreover, since $f''/f=NS^2(w^2-1)/r^2+\gamma(\varphi''+\gamma\varphi'^2)$  from \eqref{bg_3tortoise}, we get rid of the remaining singularities in $V^1$ generated by $w$,
\begin{eqnarray}
\label{V1}
V^1=\frac{NS^2}{r^2}(3w^2-1)+\gamma(\varphi''+\gamma  \varphi'^2)+2(f^2z)'.
\end{eqnarray}
To find a candidate $\psi_0$ for a zero mode of $H^1$, we look for a solution of the equation $Q^+\psi_0=0$. Using the form of the superpotential $\sigma=f'/f+f^2z$ from above, we have
\begin{eqnarray}
\label{psi0}
\psi_0=C f\,\exp \int^\rho\!\!d\rho \,\,f^2z.
\end{eqnarray}
Let's now first treat the case of the solitons. Due to the  remaining freedom in the choice of a solution of \eqref{gauge-fixing}, we construct a function $z$ in Appendix \ref{app:V1} such that both $V^1$ is sufficiently regular and $\psi_0$ is square integrable on the half-line and vanishes linearly at the origin:  Plugging \eqref{zat0} and \eqref{zatoo} into \eqref{V1} and using the asymptotics of the background from Appendix \ref{app:asymptotics}, we find that $V^1$ is a bounded $S$-wave potential with 
\begin{eqnarray}
V^1&=&C+\mO(\rho),\quad\rho\to 0,\nonumber\\
\label{V1atoo}
V^1&=& \frac{C}{\rho^2}+\mO\left(\frac{\log \rho}{\rho^3}\right),\quad\rho\to\infty.
\end{eqnarray}
Furthermore, plugging \eqref{zat0}, \eqref{zatoo} into \eqref{psi0}, the zero mode $\psi_0$ behaves like
\begin{eqnarray*}
\psi_0=C\rho+\mO(\rho^2),\quad\rho\to 0,\\
\psi_0=\mO\left(\frac{1}{\rho^2}\right),\quad \rho\to\infty,
\end{eqnarray*}
which implies the desired properties of $\psi_0$ (the non-uniqueness of $V^1$ does not affect the bound state problem, cf. \cite{V96}). Now, due to Sturm-Liouville theory \cite{AHP}, the number of bound states of $H^1$ equals the number of nodes of $\psi_0$. But, using definition \eqref{def:f,g}, the number of nodes of $\psi_0$ equals the number of nodes of $w$ because $\psi_0$ is of the form \eqref{psi0}. Since,  as long as $\omega^2<0$, the supersymmetry transformation is invertible, the claim follows.

The derivation of the number of instabilities for the black hole case proceeds along the same lines as the one just presented for the soliton case, cf.  Appendix \ref{app:BlackHole} for a brief discussion.

%%%%%%%%%%%%%%%%%%%%%%%%%%%%%%%%%%%%%%%%%%%%%%%%%%%%%%%%%%%%%%%%%%%%%%
\section{Concluding remarks}
\label{section:remarks}

We studied the linear stability of the SU(2) EYMD soliton and black hole solutions of \eqref{bg_1}-\eqref{U} under small time harmonic, spherically symmetric perturbations \eqref{perturb} with the help of the following strategy. The pulsation equations \eqref{a1}-\eqref{a0} reduce to the Schr\"odinger eigenvalue problem \eqref{H0} with Schr\"odinger operator $H^0$ whose potential $V^0$ \eqref{V0} has singularities located at the roots of the gauge amplitude $w$. Due to the residual gauge invariance \eqref{gauge-modes},\eqref{gauge-fixing}, the Hamiltonian $H^0$ has a zero mode $\phi_0$ which allows for the factorization of $H^0$ in a product of the supersymmetric charges $Q^\pm$, cf. after \eqref{V0}.  Moreover, using the gauge freedom in \eqref{gauge-fixing}, we can choose the potential $V^1$ \eqref{V1} of the supersymmetric partner Hamiltonian $H^1$ to be sufficiently regular. Since the zero mode $\psi_0$ of $H^1$ has its zeroes at the nodes of $w$ \eqref{psi0} and since the supersymmetry transformation is invertible on the negative spectrum, the Sturm-Liouville theory implies that the number of instabilities equals the number of nodes of $w$.

In \cite{BHLSV96}, for the EYM theory with a positive cosmological constant $\Lambda$, this strategy has been used to derive the linear instability of the sphaleron modes for all purely magnetic, static, spherically symmetric solutions, classified numerically in \cite{VSLHB96} and analytically in \cite{BFM05} with respect to the size of $\Lambda$ and the number $n$ of nodes of the gauge amplitude $w$ (cf. as well \cite{BFM00} for solutions with similar properties of spatially compact sections in the EYM-Higgs theory). In \cite{FR03}, a numerical analysis in the gravitational  sector revealed an unexpected dependence of the number of instabilities on the value of $\Lambda$: The number of unstable modes of the $n=3$ solution jumps from $n$ to $n-2$ as $\Lambda$ crosses a critical value from below, a phenomenon which remains to be studied in more detail.

%%%%%%%%%%%%%%%%%%%%%%%%%%%%%%%%%%%%%%%%%%%%%%%%%%%%%%%%%%%%%%%%%%%%%%
\vspace{1cm}

{\bf Acknowledgements}\,\, I'm grateful to O. Brodbeck, G. Lavrelashvili, N. Straumann, and M.S.  Volkov for many interesting discussions. Moreover, I thank the referees for their constructive remarks.

%%%%%%%%%%%%%%%%%%%%%%%%%%%%%%%%%%%%%%%%%%%%%%%%%%%%%%%%%%%%%%%%%%%%%%
\begin{appendix}

\section{Constructing $z$}
\label{app:V1}

Using the definitions \eqref{def:x,y}, \eqref{def:f,g} and the background equation \eqref{bg_1}, we can rewrite \eqref{gauge-fixing} as a second order linear differential equation of Fuchsian type in the variable $r$,
\begin{eqnarray}
\label{h''}
r^2h''+2r\left(1\!-\!w'^2\!+\!\frac{r^2}{2}\varphi'^2\!+\!\gamma r\varphi'\right)h'-\frac{2w^2}{N}h=0.
\end{eqnarray}
This differential equation has two regular singular points, a first one at the origin, and a second one at infinity. As in \cite{LM93}, we assume all the background fields to be analytic at the origin and at infinity (in the EYM case this has been proven in \cite{SW93}). Hence, using the structure of the coefficients in \eqref{h''} and the asymptotics from Appendix \ref{app:asymptotics}, the Fuchsian theory yields the indices $1$ and $-2$ at the origin,  and $2$ and $-1$ at infinity. Moreover, it turns out that the solution at the origin neither contains a $1/r$ nor a $\log r$ term in its power series expansion.  If we choose a special solution $h_0$ of \eqref{gauge-fixing} which vanishes at infinity, we have to keep the $1/r^2$ term at the origin though, cf. \cite{V96}. Since, with \cite{LM93},  we assume continuity of the coefficients in \eqref{h''}, we are thus given a classical solution of \eqref{h''} on the half-line with asymptotics
\begin{eqnarray}
\label{h0at0}
h_0&=&\frac{C}{r^2}+\mO(1),\quad r\to0,\\
\label{h0atoo}
h_0&=&\frac{C}{r^2}+\mO\left(\frac{1}{r^3}\right),\quad r\to\infty.
\end{eqnarray}
To pick a function $z$ having the desired properties described after \eqref{psi0}, we make use  of the general solution of \eqref{gauge-fixing} constructed from the particular solution $h_0$. In the  coordinate $\rho$, the general $z$ takes the form
\begin{eqnarray}
\label{z}
z=-h_0\,\frac{g}{h_0'}-\left(\frac{g}{h_0'}\right)^2\left(C\!+\!\int^\rho\!\!\!d\rho\, \,f^2\left(\frac{g}{h_0'}\right)^2\right)^{-1}.
\end{eqnarray}
Plugging the expressions for the asymptotic behavior from Appendix  \ref{app:asymptotics} and \eqref{h0at0},\eqref{h0atoo} into \eqref{z}, we can achieve that
\begin{eqnarray}
\label{zat0}
z&=&\frac{e^{-2\gamma\varphi_0}}{\rho}+\mO(\rho),\quad\rho\to 0,\\
\label{zatoo}
z&=&-\frac{2e^{-2\gamma\varphi_\infty}}{\rho}+\mO\left(\frac{\log \rho}{\rho^2}\right),\quad\rho\to\infty.
\end{eqnarray}

%%%%%%%%%%%%%%%%%%%%%%%%%%%%%%%%%%%%%%%%%%%%%%%%%%%%%%%%%%%%%%%%%%%%%%
\section{Asymptotics of the background}
\label{app:asymptotics}

From \cite{LM93} we have the following asymptotic behavior of the fields at the origin (note that $S$ can be expressed through $w$ and $\varphi$ with the help of the background equation \eqref{bg_1}),
\begin{eqnarray*}
\varphi&=&\varphi_0+2\gamma e^{2\gamma\varphi_0}b^2r^2+\mO(r^4),\\
w&=&1-br^2+\mO(r^4),\\
N&=&1-4e^{2\gamma\varphi_0}b^2r^2+\mO(r^4),\\
S&=&1+4e^{2\gamma\varphi_0}b^2r^2+\mO(r^4).
\end{eqnarray*}
At infinity, the asymptotics looks like
\begin{eqnarray*}
\varphi&=&\varphi_\infty-\frac{D}{r}+\mO\left(\frac{1}{r^2}\right),\\
w&=&\pm\left(1-\frac{c}{r}\right)+\mO\left(\frac{1}{r^2}\right),\\
N&=&1-\frac{2M}{r}+\frac{D^2}{r^2}+\mO\left(\frac{1}{r^3}\right),\\
S&=&S_\infty\left(1-\frac{D^2}{2r^2}\right)+\mO\left(\frac{1}{r^4}\right),
\end{eqnarray*}
where $M$ is the ADM mass and $D$ stands for the dilaton charge. For a more detailed description and the meaning of the other constants, cf. \cite{LM93}.

%%%%%%%%%%%%%%%%%%%%%%%%%%%%%%%%%%%%%%%%%%%%%%%%%%%%%%%%%%%%%%%%%%%%%%
\section{The black hole case}
\label{app:BlackHole}

The starting point is again equation \eqref{gauge-fixing}. In the black hole case, the coordinate $\rho$ from \eqref{def:tortoise} runs over the whole real line with $\rho=-\infty$ at the horizon, $r_h$ say. The two regular singular points are now at $\rho=\pm\infty$. Rewriting equation \eqref{h''} in the coordinate $q=(r-r_h)/r_h$, assuming analyticity of the background solutions in some neighborhood of the horizon $r_h$ (cf. \cite{SW93} for the EYM case), and using the asymptotics \cite{LM93}
\begin{eqnarray*}
\varphi&=&\varphi_h-\frac{d}{r}+\gamma\frac{V_h}{1-V_h}q+\mO\left(q^2\right),\\
w&=&w_h+\frac{w_h\left(w_h^2-1\right)}{1-V_h}q+\mO\left(q^2\right),\\
N&=&(1-V_h)q+\mO\left(q^2\right),\\
S&=&S_h\left(1+\frac{(2w_h^2+\gamma^2V_h)V_h}{(1-V_h)^2}q\right)+\mO\left(q^2\right),
\end{eqnarray*}
where $V_h=e^{2\gamma\varphi_h}(w_h^2-1)^2/r_h^2$, the Fuchsian theory yields the indices $1$ and $0$ (note that $1-V_h>0$). The asymptotics at infinity is described in Appendix \ref{app:asymptotics} with the indices from Appendix \ref{app:V1}. Choosing the particular solution $h_0$ of \eqref{gauge-fixing} to be the power series  belonging to index $1$, we find, in the same way as in Appendix \ref{app:V1}, that
\begin{eqnarray*}
h_0&=&Cq+\mO(q^2),\quad q\to 0,\\
h_0&=&Cq+\mO(1),\quad q\to\infty.
\end{eqnarray*}
Hence, with \eqref{z}, the general $z$ has the form \eqref{zatoo} for $\rho\to\infty$, and
\begin{eqnarray*}
z=-\frac{e^{-2\gamma\varphi_h}}{w_h^2\rho}+\mO\left(\frac{1}{\rho^2}\right),\quad \rho\to-\infty.
\end{eqnarray*}
Furthermore, the potential $V^1$ decays as $1/\rho^2$ in leading order both for $\rho\to\infty$ and $\rho\to-\infty$ (cf. \eqref{V1atoo}), and the zero mode $\psi_0$ from \eqref{psi0} is square integrable on the whole real line,
\begin{eqnarray*}
\psi_0&=&\mO\left(\frac{1}{|\rho|}\right),\quad\rho\to -\infty,\\
\psi_0&=&\mO\left(\frac{1}{\rho^2}\right),\quad\rho\to \infty.
\end{eqnarray*}

\end{appendix}

%%%%%%%%%%%%%%%%%%%%%%%%%%%%%%%%%%%%%%%%%%%%%%%%%%%%%%%%%%%%%%%%%%%%%%

%%%%%%%%%%%%%%%%%%%%%%%%%%%%%%%%%%%%%%%%%%%%%%%%%%%%%%%%%%%%%%%%%%%%%%

\end{document}